\documentclass[12pt, preprint]{aastex}
\usepackage{graphicx}	
\usepackage{natbib}	
\usepackage{amsmath}	
\usepackage{bbm}	
\usepackage[breaklinks]{hyperref}

\newcommand{\sex}{{\tt SExtractor}}

\shorttitle{Probabilistic Stellar Catalogs}
\shortauthors{Brewer et al.}

\begin{document}

\title{Probabilistic Catalogs for Crowded Stellar Fields}

\author{Brendon J. Brewer\thanks{\texttt{bj.brewer@auckland.ac.nz}}}
\affil{Department of Statistics, The University of Auckland,
Private Bag 92019, Auckland 1142, New Zealand}


\author{Daniel Foreman-Mackey}
\affil{Center for Cosmology and Particle Physics, Department of Physics,
New York University, Washington Place, New York, NY, 10003, USA}

\and

\author{David W. Hogg}
\affil{Center for Cosmology and Particle Physics, Department of Physics,
New York University, Washington Place, New York, NY, 10003, USA}

\begin{abstract}
We present and implement a probabilistic (Bayesian) method for producing catalogs
from images of stellar fields. The method is capable of inferring the
number of sources $N$ in the image and can also handle the
challenges introduced by noise, overlapping sources, and an unknown
point spread function (PSF). The luminosity
function of the stars can also be inferred even when the precise luminosity
of each star is uncertain, via the use of a hierarchical Bayesian model.
The computational feasibility of the method is
demonstrated on two simulated images with different numbers of stars.
We find that our method
successfully recovers the input parameter values along with principled
uncertainties even when the field is crowded.
We also compare our results with those obtained from the
\sex~software. While the two approaches largely agree about the fluxes of the
bright stars, the Bayesian approach provides more accurate inferences about
the faint stars and the number of stars, particularly in the crowded case.
\end{abstract}


\section{Introduction}
Traditional practice in astronomy is to take images of the sky, detect
or enumerate sources visible in those images, and create catalogs.
These catalogs are then used to perform fundamental astronomical
measurements, for example reconstructing the three-dimensional
structure of the Galaxy or the two-point correlation function of
galaxies.  Indeed, the process of catalog construction is so ``baked
in'' to our ideas about what astronomy is, we sometimes forget that
the catalog is \emph{not} the fundamental data product of astronomy; catalogs are produced
from imaging; their production involves many decisions and ideas
that go beyond the information provided to the telescope by the
incident intensity field. In addition, catalogs are not usually the final
goal of any imaging project or survey. Typically, they are produced in
order to facilitate the scientific study of populations of objects (e.g. the
initial mass function of a population of stars), or to provide a sky-search
capability to the community who might be interested in only a small subset of
objects. Standard tools for generating catalogs from astronomical imaging
include \sex~\citep{sextractor}, DAOPHOT \citep{1987PASP...99..191S}, 
DOLPHOT \citep{dolphot}, and SDSS
Photo \citep{photo}.

Telescopes don't make catalogs \citep{2011EAS....45..351H}, they
measure the intensity field.  Viewed through the lens of probabilistic
inference, the goals of astronomy are to take the information in the
telescope-generated records of the intensity field and use this information
to obtain quantities of astronomical interest with as little loss as
possible.  Insertion of a catalog-generation step in the inference
pipeline between the raw imaging and the final astrophysical analyses
is potentially lossy. The hard decisions of catalog making destroy
information, at least in principle.  Probability theory suggests that it may be less
lossy to pass forward not a catalog but a probabilisitic
description of all the catalogs that could be consistent with the
imaging---a posterior probability distribution in the (enormously large) space of possible catalogs.
Essentially, the creation of a catalog is an
attempt to answer the question,
``Given the image we have obtained, what objects are present in the field and
what are their properties?''.
This article represents an attempt at implementing this ambitous goal in the specific
situation where the only objects in the field are stars or other point sources.

Beyond these philosophical concerns, there are practical issues;
standard methods for constructing catalogs can have difficulty in some
challenging situations. For example, when multiple sources overlap partially
or completely, it can be difficult to determine how many sources are present,
and how much flux belongs to each source. In principle, the uncertainty about
the existence and properties of the objects can be significant and should be
propagated into any inferences about the stellar population. A Bayesian approach
that obtains the posterior distribution over catalog space (rather than a single
catalog estimate) has the potential to overcome these problems by deblending
objects when it is possible, and clearly indicating the uncertainty remaining
when it is not possible.

In practice, Bayesian Inferences are often implemented using Markov Chain Monte
Carlo (MCMC) methods \citep{mackay} to sample from the posterior distribution.
Sampling a posterior probability distribution for catalogs is a challenging
numerical task for a number of reasons. Firstly, the number $N$ of objects in the
image (and that should therefore be listed in the catalog) is itself unknown.
Secondly, if $N$ is large, then the parameter space
of positions and properties (flux, size, etc) of the objects is also large.
This can cause Markov Chain Monte Carlo (MCMC) algorithms difficulties -- they
may take a long time to converge to the target posterior distribution over
the space of catalogs. Thirdly, this problem is subject to the so-called
label-switching problem that is commonly encountered in mixture modeling
\citep[e.g.][]{label_switching}. Given any proposed catalog, another catalog that is
equally plausible is the catalog obtained by shuffling the entries of the first
catalog. This leads to a posterior distribution with $N!$ identical peaks in
parameter space. This can lead to difficulties with certain (otherwise very
effective) MCMC algorithms such as the affine-invariant stretch move
\citep{goodman, emcee}.

Bayesian object detection (as this problem is sometimes called) has been
implemented both inside and outside of astronomy
\citep[e.g.][]{object, 2003MNRAS.338..765H, 2011MNRAS.415.3462F}. However, the
\citet{2011MNRAS.415.3462F} approach makes the assumption of a
known number of objects $N$.
This assumption is required for the
{\tt MultiNest} sampler \citep{multinest} to be applicable.
Using the results from the known $N$
run, it is possible (under certain circumstances) to reconstruct what the
results would have been if an unknown-$N$ model had been used. However,
this will not work well in situations where there is significant confusion
(i.e. two or more sources overlap). What is really required is a variable
dimension model, where $N \in \{0, 1, 2, ... \}$ is an unknown quantity to be
inferred from the data \citep[e.g.][]{2003MNRAS.338..765H}.
The computational implementation of these models will require tools such as
reversible jump Markov Chain Monte Carlo \citep{rjmcmc}. Other statistical
methods have also been used to model crowded fields
\citep[e.g. maximum likelihood,][]{irwin}.
However, maximum likelihood is not completely appropriate in flexible models
because it may lead to overfitting. In this situation overfitting would result
in more stars being added to the model to explain small positive fluctuations
in the image which are actually due to noise. Various other techniques have
also been proposed in the literature
\citep[e.g.][]{2002MNRAS.335...73M, 2007A&A...461..373M, 2009PASP..121.1429Z}.

In this paper, we develop a Bayesian object detection model with the following
features: i) the number $N$ of objects in the image is an unknown parameter
to be inferred from the data, ii) the objects that we expect to find are point sources such
as stars, and iii) the point-spread function is unknown (but a parametric model is
used) and must be inferred from
the data (but a single bright star may not be available to help with estimating it).
The paper is structured as follows. In Section~\ref{sec:bayes}
we give a brief overview of Bayesian inference. In Section~\ref{sec:model} we
discuss the model assumptions we make in our method. In Section~\ref{sec:mcmc}
we briefly discuss our
MCMC implementation. Section~\ref{sec:simulated_data} describes the tests we
carried out on simulated data, and a comparison with \sex~results is presented
in Section~\ref{sec:sex}. We conclude in Section~\ref{sec:conclusion}.

\section{Bayesian Inference}\label{sec:bayes}
To quantitatively model uncertainties and transform noise in observed data
into uncertainties in parameters of interest, Bayesian Inference is the appropriate
framework \citep{cox, jaynes, caticha, mackay}. Suppose there exist unknown parameters
(denoted
collectively by $\theta$) and we expect to obtain some data $x$. Our prior
state of knowledge about the parameters is modelled by a prior
probability distribution:
\begin{equation}
p(\theta).
\end{equation}
Note that this is a very concise notation \citep{hogg} and should be read
as, ``The probability distribution for $\theta$''.
We also model how the parameters give rise to the data, via a generative model.
This is also known as a {\it sampling distribution}:
\begin{equation}
p(x|\theta).
\end{equation}
Despite the singular, the sampling distribution is actually a family of
probability distributions over the space of possible data sets, one probability
distribution for each possible value
of $\theta$. Note that the choice of
the sampling distribution is also an assumption about prior knowledge:
It models prior information about the fact that the data $x$ is connected to
the parameters $\theta$ in some way \citep{caticha}. Without this prior
knowledge, learning is impossible: there has to be some relationship between
the parameters and the data, otherwise it would be impossible to learn about
parameters by obtaining data.

When specific data $x^*$ are taken into account, our state of knowledge about $\theta$
gets updated from the prior distribution to the posterior distribution
via Bayes' rule:
\begin{eqnarray}
p(\theta|x=x^*) &\propto& p(\theta)p(x|\theta)|_{x=x^*} \\
&=& p(\theta)\mathcal{L}(\theta; x)
\end{eqnarray}
The term $p(x|\theta)|_{x=x^*} = \mathcal{L}(\theta; x)$
is the {\it likelihood function}, which is the
probability of obtaining the actual data set $x^*$ as a function of the
parameters. In the case that the sampling distribution is a probability
density function, the likelihood is the probability density function evaluated at the observed
data. This usually causes no problems, although one should be aware of the
Borel-Kolmogorov paradox \citep{jaynes}.
As suggested by the above notation, the likelihood function is obtained from the
sampling distribution with the actual data substituted in and is therefore
a function of the parameters only.

To proceed with the model for inferring catalogs
from image data, we must
specify a definite hypothesis space and choices for the prior distribution
and the sampling distribution. These choices are presented and discussed in
Section~\ref{sec:model}.

\section{The Specific Model for Stellar Fields}\label{sec:model}
\subsection{The Hypothesis Space}
The hypothesis space is the set of possible catalogs, or the set of possible
answers to the question, ``What objects are present in the field and
what are their properties?'' We shall assume that there are an unknown number of stars
$N$ in the field. Each star has an unknown
position $(x,y)$ in the plane of the sky, and an unknown flux $f$. We also
describe the distribution of fluxes (commonly known as the {\it luminosity
function}) of the stars by some parameters denoted collectively by $\beta$.
In summary, the unknown parameters are:
\begin{eqnarray}
\theta = \left\{N, \beta, \left\{x_i, y_i\right\}_{i=1}^N,
\left\{f_i\right\}_{i=1}^N\right\}.
\end{eqnarray}
We note that models similar to this have been implemented for general image
modeling and deconvolution \citep[e.g.][]{massinf}, however in this case it is
more justified as we are actually searching for point fluxes.

\subsection{The Prior}
The prior probability distribution for the unknown parameters can be factorized
using the product rule of probability theory.
With a variety of independence assumptions, the prior
can be factorized as:
\begin{eqnarray}
p(\theta) = p(\beta)p(N|\beta)\prod_{i=1}^N p(x_i, y_i)
p(f_i | \beta)
\end{eqnarray}
Here, we have assumed that the luminosity function does not depend on position.
Finally, the fluxes of the stars come independently from
a common distribution. If we knew the luminosity
function of the stars, then the location and flux of a particular star would
not tell us anything about the location and flux of another star. Really, this
is just a way of implementing exchangeability of the stars, and is often
called a {\it hierarchical model}.

For simplicity, we assume a uniform prior probability distribution for the
position of each star. The use of independent priors for the positions creates
a strong preference for catalogs where
the stars are uniformly scattered across the image. Thus, this model is
appropriate for small images where the density of stars is approximately
uniform across the image. In other scenarios, such as images of stellar clusters, it is possible
to parameterize the spatial distribution of the stars in a similar way to how
we have parameterized the luminosity function, i.e. as a hierarchical model.

\subsection{The Sampling Distribution}
The sampling distribution is a probabilistic model for the process that
generates the data; it describes the probability distribution we would use
to predict the data if we happened to know the true catalog. In our case,
the data will be an $m \times n$ array of pixel intensities $I$:
\begin{eqnarray}
\{I_{ij}\}
\end{eqnarray}
where the central position of each pixel is:
\begin{eqnarray}
\{X_{ij}, Y_{ij}\}.
\end{eqnarray}
The image is assumed to be a noisy version of the true underlying intensity
field. Thus, we need
a prescription for simulating an image $\{I_{ij}\}$ from a catalog $\theta$:
\begin{eqnarray}
\theta = \left\{N, \beta, \left\{x_i, y_i\right\}_{i=1}^N,
\left\{f_i\right\}_{i=1}^N\right\}.
\end{eqnarray}
If we knew the true catalog, we could compute the
``mock image'' we would expect to see
in the absence of noise. This mock image (defined at every point on the sky)
is given by:
\begin{eqnarray}
\mathcal{M}(x, y) &=& \sum_{i=1}^N f_i \mathcal{P}(x - x_i, y - y_i)
\end{eqnarray}
where $\mathcal{P}$ is the pixel-convolved point spread function (PSF). The use
of a pixel-convolved PSF is computationally advantageous because the PSF need
only be evaluated at the center of each pixel, rather than integrated over each
pixel.

Throughout this paper we will
assume the pixel-convolved PSF is a weighted mixture of two concentric
circular
Gaussians (a similar mixture model that uses a Gaussian core and flexible wings is also used by DAOPHOT, \citet{1987PASP...99..191S}) with widths
$s_1$ and $s_2$:
\begin{eqnarray}
\mathcal{P}(x, y) = \frac{w}{2\pi s_1^2}\exp
\left[
-\frac{1}{2s_1^2}\left(x^2 + y^2\right)
\right]
+ \frac{1-w}{2\pi s_2^2}\exp
\left[-\frac{1}{2s_2^2}\left(x^2 + y^2\right)
\right].
\end{eqnarray}
The pixellated observed image is assumed to be generated from the mock
image (evaluated at the center of each pixel) plus noise:
\begin{eqnarray}
I_{ij} &=& \mathcal{M}(X_{ij}, Y_{ij}) + \epsilon_{ij}
\end{eqnarray}
where the errors $\{\epsilon_{ij}\}$ are independent and normally distributed.
The variance of the normal distribution for each pixel is determined by the
brightness of the sky and the
brightness of the mock image in that pixel. This can be modelled by assuming the
following distribution:
\begin{eqnarray}
\epsilon_{ij} \sim \mathcal{N}(0, \sigma_0^2 + \eta \mathcal{M}(X_{ij}, Y_{ij})).
\end{eqnarray}
where $\sigma_0$ is a constant noise level and $\eta$ is an unknown = coefficient that allows for the possibility that
the noise level is higher in brighter regions of the image. This dependence of the
noise variances on the model intensity arises as a result of the Poissonian
nature of photon counts, but allows for the fact that a ``sky'' background may
have already been subtracted from the image in the reduction process.
This parameterisation has been used by \citet{2011MNRAS.412.2521B} and is an
alternative to the common practice of producing a ``variance map'' from the
image data that is then assumed to be known.

\subsection{The Prior Distribution}
The prior distribution for the number of stars $N$ is assigned to be uniform
between 0 and some maximum number $N_{\rm max}$. The extent of the image is
assumed to be from $x=x_{\rm min}$ to $x=x_{\rm max}=x_{\rm min} + x_{\rm range}$ and from
$y=y_{\rm min}$ to $y=y_{\rm max}=y_{\rm min} + y_{\rm range}$ in arbitrary units, and the
positions of the stars are assigned independent uniform priors:
\begin{eqnarray}
x_i &\sim& \textnormal{Uniform}(x_{\rm min} - 0.1x_{\rm range}, x_{\rm max} + 0.1x_{\rm range}) \\
y_i &\sim& \textnormal{Uniform}(y_{\rm min} - 0.1y_{\rm range}, y_{\rm max} + 0.1y_{\rm range})
\end{eqnarray}
The stars are allowed to be slightly outside of the observed image because the
PSF can scatter light from these stars into the image.

For the purposes of this paper, we model the luminosity function as a broken
power-law distribution, which has four
free parameters:
\begin{eqnarray}
\beta = \{h_1, h_2, \alpha_1, \alpha_2\}.
\end{eqnarray}
where $h_1$ is a lower flux limit, $h_2$ is a break-point, $\alpha_1$ is
the slope of the distribution between $h_1$ and $h_2$, and $\alpha_2$ is the
slope of the distribution above $h_2$. For mathematical details on the broken power-law
model, see Appendix~\ref{power_law}. While the broken power-law is likely to be
unrealistic in many cases, it is a reasonably flexible distribution and this is
sufficient for demonstrating the properties of our method.

The prior distribution on $h_1$, $h_2$,
$\alpha_1$, and $\alpha_2$ is assigned to be:
\begin{eqnarray}
\ln h_1 &\sim& \textnormal{Uniform}(\ln(10^{-3}), \ln(10^3)) \\
\ln h_2 &\sim& \textnormal{Uniform}(\ln(h_1), \ln(h_1) + 2.3) \\
\alpha_1 &\sim& \textnormal{Uniform}(1, 5) \\
\alpha_2 &\sim& \textnormal{Uniform}(1, 5).
\end{eqnarray}
These priors express vague prior knowledge about $\alpha_1$ and $\alpha_2$ in
addition to vague prior knowledge about $h_1$ and $h_2$ apart from the fact
that the flux units are not extreme and that $h_2$ should be no more than an
order of magnitude greater than $h_1$.

This simply-parameterized model for the luminosity function can be criticized
on the basis that information from bright stars can be used to infer the
parameters of the luminosity function which then still apply at lower flux
levels. In principle, this can be resolved by using a more flexible distribution
\citep[e.g.][]{2008ApJ...682..874K} where each star's measured brightness
affects the inference of the luminosity function locally but not globally.

The priors for the PSF parameters and the noise parameters were assigned to be:
\begin{eqnarray}
\ln s_1 &\sim& \textnormal{Uniform}(\ln(0.3L), \ln(30L)) \\
\ln s_2 &\sim& \textnormal{Uniform}(\ln(s_1), \ln(s_1) + 2.3) \\
w &\sim& \textnormal{Uniform}(0, 1)\\
\ln \sigma_0 &\sim& \textnormal{Uniform}(\ln(10^{-3}), \ln(10^{3})) \\
\ln \eta &\sim& \textnormal{Uniform}(\ln(10^{-3}), \ln(10^{3}))
\end{eqnarray}
where $L = x_{\rm range}/n$ is the width of a pixel. These priors
describe vague prior knowledge about the overall scale of the PSF except that the
wider component is less than 10 times as wide as the narrow component, as well
as the knowledge that the noise variance is not extreme relative to the fluxes
of the stars.

\section{MCMC Implementation}\label{sec:mcmc}
The MCMC sampling was implemented using the Diffusive Nested Sampling
\citep{dnest} method (hereafter DNS). DNS is a variant of the
Nested Sampling \citep{skilling} algorithm that uses Metropolis-Hastings
updates, and
is very generally applicable. The main difference between DNS and the standard
Metropolis-Hastings algorithm is that the target distribution is modified.
Rather than simply exploring the posterior distribution over catalog space,
DNS constructs an alternative target distribution which is a mixture of the
prior distribution with more constrained versions of the prior distribution.
The modified target distribution assists the sampling in several ways.
Firstly, the target distribution shrinks at a constant rate with time during the
initial phase of the exploration. This is similar to the popular ``simulated
annealing'' method \citep{annealing, neal} but with an optimal annealing schedule. Secondly,
communication with the prior is maintained: once a catalog is found that
fits the data, the catalog can ``disintegrate'' back to the prior distribution
and re-fit, allowing different peaks in the parameter space to be explored
(if they exist). This all happens naturally within the context of a valid
MCMC sampler. The MCMC may also be run using the standard Metropolis algorithm
targeting the posterior distribution.

\section{Simulated Data}\label{sec:simulated_data}
\begin{figure}[ht!]
\begin{center}
\includegraphics[width=\textwidth]{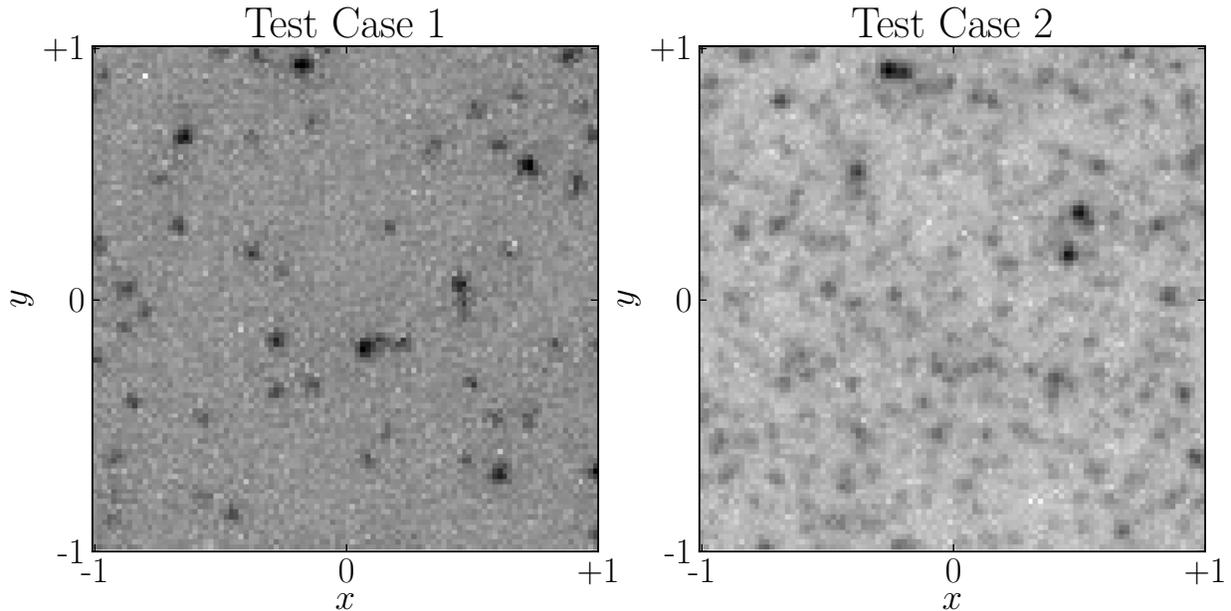}
\caption{The two simulated images used to test our methodology.
{\bf Left}: An image containing $\sim$100 stars.
{\bf Right}: An image containing $\sim$1000 stars.\label{fig:simulated_data}}
\end{center}
\end{figure}

In order to test our approach, we applied the method to two illustrative
simulated images generated from
the above model (Figure~\ref{fig:simulated_data}). The purpose of this
experiment was to test the computational
feasibility of the model, as well as to compare the inferences from the model
with those from more standard techniques.

The true parameter values for the two simulated data sets are listed in
Table~\ref{tab:truth}. The broken power-law parameter values were chosen so that
roughly half of the stars' fluxes were below and above the break-point
respectively. Figure~\ref{fig:powerlaw} in Appendix~\ref{power_law} also shows the true flux distribution
used for the simulated images. Each of the images
is $100 \times 100$ pixels in extent and covers a range from $-1$ to $1$ in
arbitrary units for both the $x$ and $y$ axes. The first image contains
100 stars (including stars just outside of the image; there are 63 stars whose central
positions lie within the image) and the second
image contains $\sim$1000 stars (699 of which are positioned within the image).

\begin{table}[ht!]\footnotesize
\begin{center}
\begin{tabular}{|c|c|c|}
\hline
Parameter & Value (Test Case 1) & Value (Test Case 2)\\
\hline
$N$ & 100 & 1000\\
$h_1$ & 0.3 & 0.3\\
$h_2$ & 0.6 & 0.6\\
$\alpha_1$ & 1.1 & 1.1\\
$\alpha_2$ & 2 & 2\\
\hline
$\sigma_0$ & 10 & 10\\
$\eta$ & 10 & 10 \\
$s_1$ & 0.02 & 0.02\\
$s_2$ & 0.1 & 0.1\\
$w$ & 0.5 & 0.5\\
\hline
\end{tabular}
\end{center}
\caption{True parameter values used to generate the simulated data. 
$N$ is the number of stars, $h_1$ and $h_2$ are the lower limit and break
point of the flux distribution respectively, and $\alpha_1$ and $\alpha_2$ 
are the slopes of the flux distribution. $\sigma_0$ and $\eta$ describe the
noise properties, and $s_1$, $s_2$, and $w$ are the PSF parameters. The only
difference between the two images is that Test Case 2 contains more stars than
Test Case 1.
\label{tab:truth}}
\end{table}

\subsection{Test Case 1}
\begin{figure}[ht!]
\begin{center}
\includegraphics[width=\textwidth]{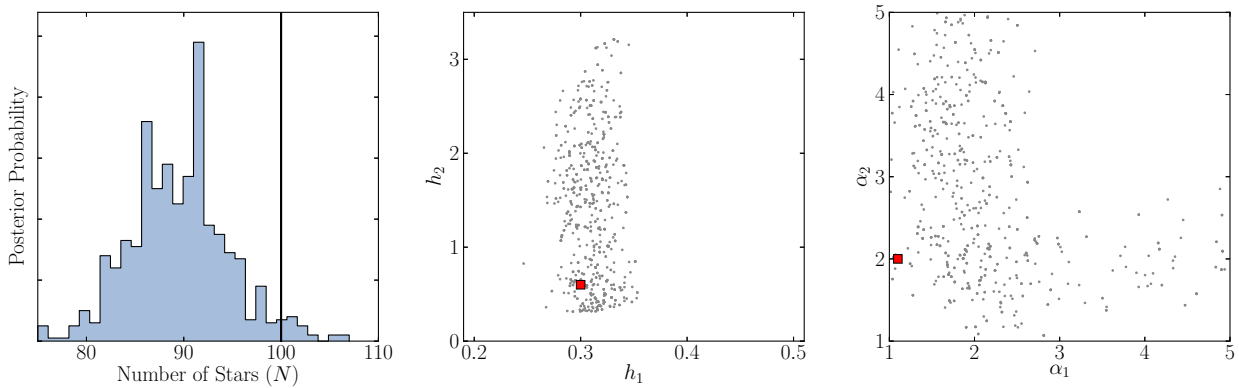}
\end{center}
\caption{Inference about the parameters for Test Case 1. The left panel shows
the posterior distribution for the number of stars $N$, and the right
panels show the joint posterior distributions for the flux distribution
parameters. Note that there is
considerable uncertainty (particularly about $h_2$), which occurs because the
shape of the broken power-law does not depend strongly on the parameters.
The true input values are plotted as filled squares.
\label{fig:results1}}
\end{figure}

Test Case 1 was run with the DNS algorithm and usable results were obtained
within about an hour on a modern desktop PC. The inferences on the parameters
$N$, $h_1$, $h_2$, $\alpha_1$, and $\alpha_2$ are shown in
Figure~\ref{fig:results1}. The number of stars is correctly inferred, and the
posterior distributions for the other parameters comfortably contain the true
input values. As $N$ is a parameter of our model, there is no need for Bayesian
``Model Comparison'' calculations to be done between different values of $N$.
The DNS method does compute the ``evidence'' value that is required for model
comparison, but this is useful only to test completely separate models, it
is not needed to infer the value of $N$.

Note that the uncertainty in $h_2$, $\alpha_1$, and $\alpha_2$ is
quite large. This is because the broken power-law model
(Figure~\ref{fig:powerlaw}) does not change drastically in shape as the
parameters are varied. Therefore, a large number of stars would be required to
tightly constrain these parameters.

The PSF parameters $\{s_1, s_2, w\}$ and the noise parameters $\{\sigma_0, \eta\}$
were also inferred accurately with small uncertainties.

\subsection{Test Case 2}
Test Case 2 is more challenging than Test Case 1 because the image contains
more stars. This increases the size of the computational task in two ways:
firstly, there will be more unknown parameters to infer, so any MCMC
algorithm will require more iterations in order to converge to the posterior
distribution. Secondly, the time taken
to compute the predicted image from a proposed catalog (in order to evaluate the
likelihood) is longer because of the larger number of stars. Hence, each MCMC
step also takes more time. Using DNS, some
samples from the posterior distribution can be obtained in about a day on a
modern multi-core PC.

\begin{figure}[ht!]
\begin{center}
\includegraphics[width=\textwidth]{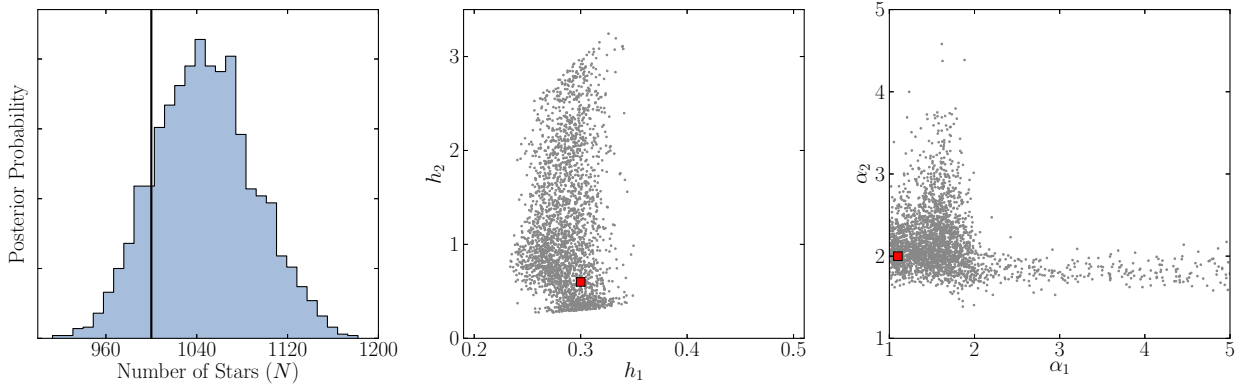}
\end{center}
\caption{Inference about the parameters for Test Case 2. Note that the
flux distribution parameters are still not very well constrained even with
the larger number of
stars. This occurs because the fluxes of faint stars are not accurately measured
and because the shape of the broken power-law distribution does not vary rapidly
as a function of its parameters. The true input values are plotted as filled squares.\label{fig:results2}}
\end{figure}

\begin{figure}[ht!]
\begin{center}
\includegraphics[scale=0.6]{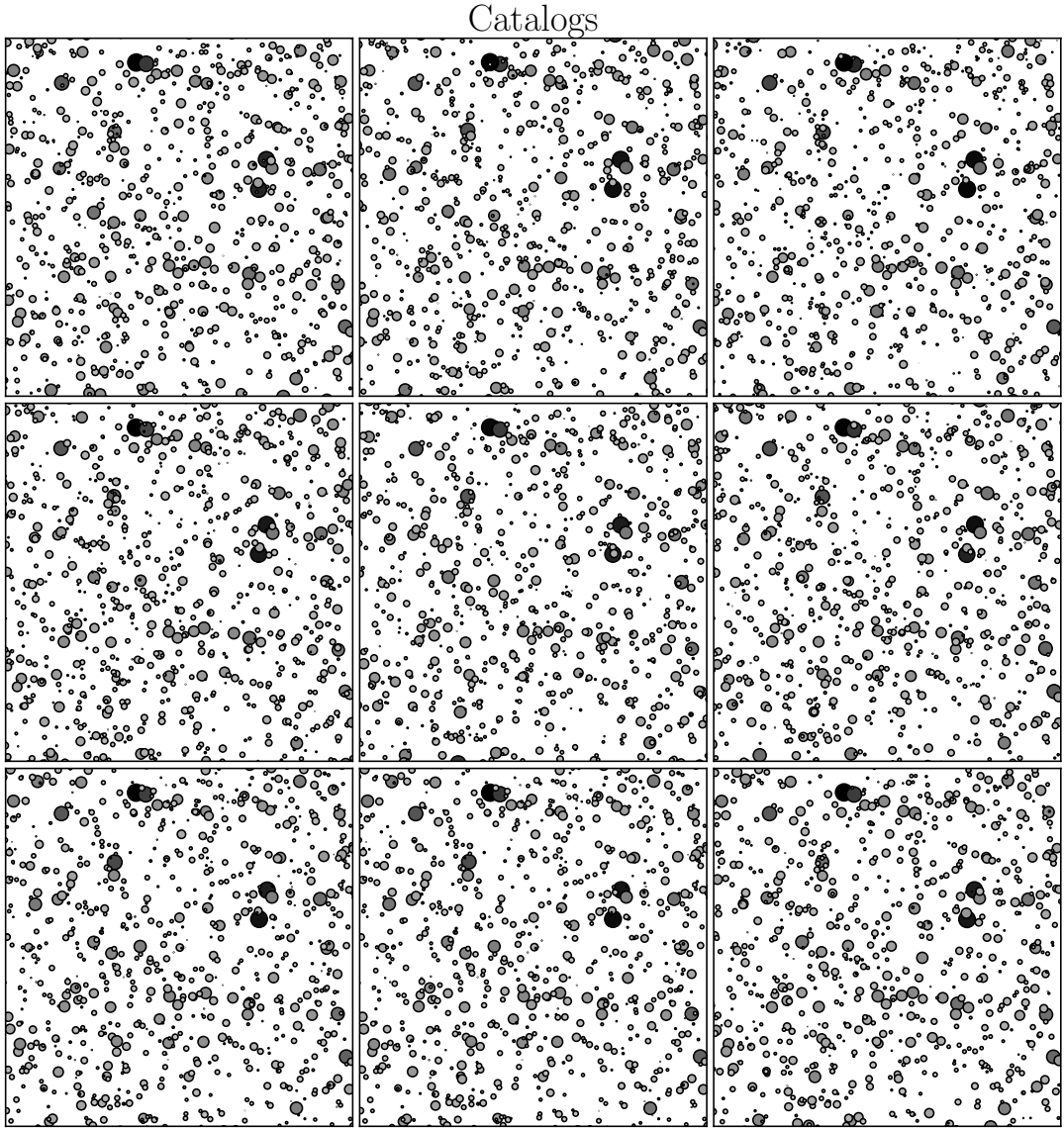}
\end{center}
\caption{Fundamentally, the output from our method is samples from the posterior
distribution over the catalog space. Nine example catalogs are shown, sampled from the posterior distribution for
Test Case 2. Features in common represent features with high probability,
and differences between the catalogs represent conclusions that are uncertain.
The area of each circle is proportional to the flux of the star.
The posterior samples may be used to compute summary images; some these are presented
in Figure~\ref{fig:summaries}.
\label{fig:catalogs}}
\end{figure}

Each catalog in the posterior sample represents a scenario for the true
underlying image that we would observe if we had a hypothetical noise-free,
infinite resolution telescope. Figure~\ref{fig:catalogs} shows nine possible catalogs
sampled from the posterior distribution. Features that are common to these nine samples are plausible, and features that differ are uncertain.

\begin{figure}[ht!]
\begin{center}
\includegraphics[width=\textwidth]{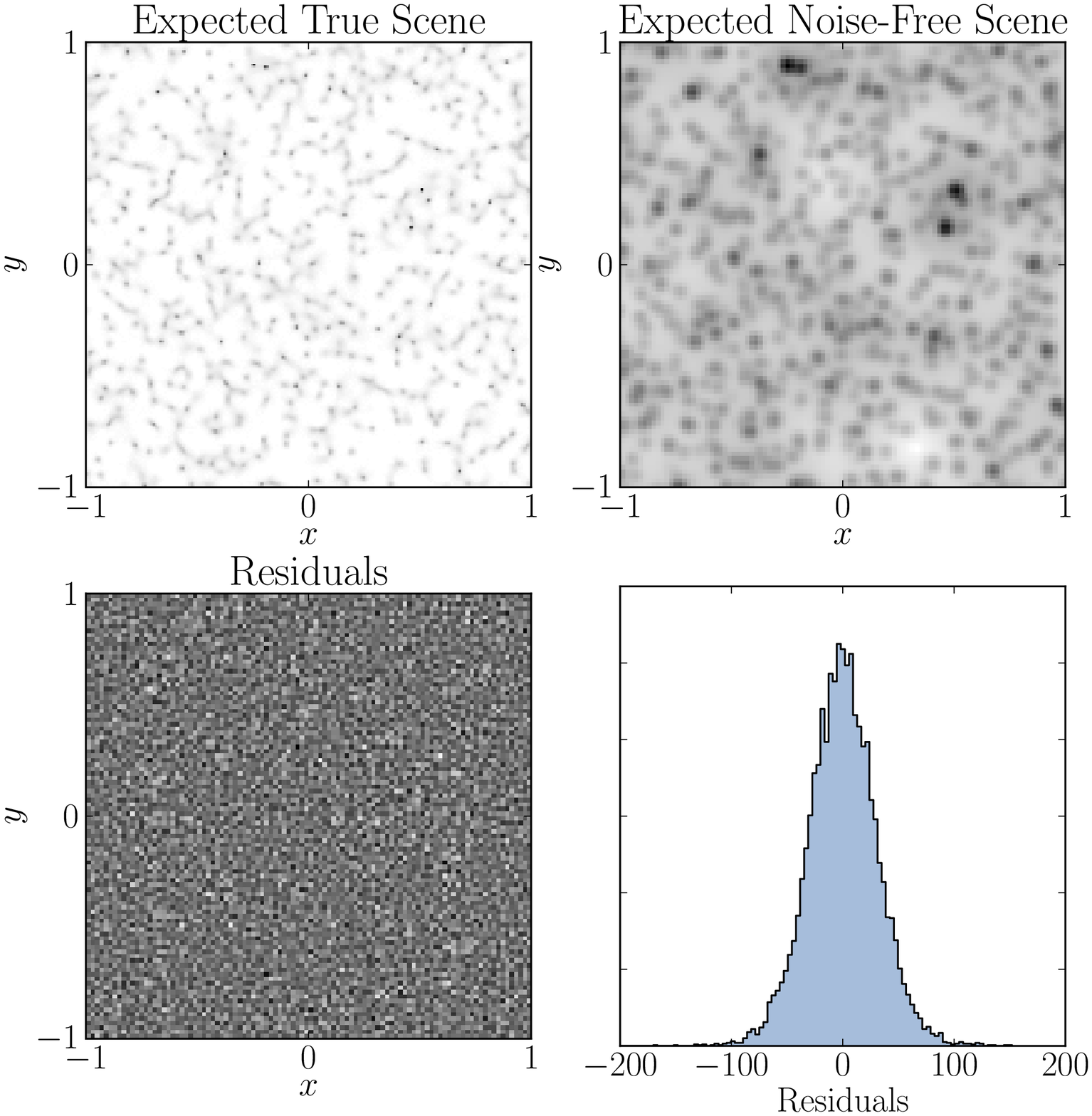}
\end{center}
\caption{Summary images produced from the posterior distribution for Test Case 2. The upper left panel shows the posterior
mean high-resolution scene. The upper right panel shows the posterior mean scene
when observed at the resolution of the data, and the bottom panels show the
standardized model residuals.
\label{fig:summaries}}
\end{figure}

From these samples, we can construct the
posterior expected true scene and other summaries. Summary images are shown in Figure~\ref{fig:summaries}. The residuals
provide a check on the validity of the model assumptions, and the posterior expected true
scene provides a useful visual guide to the uncertainties present in the catalogs. In this example,
the residuals show only noise because the simulated images were actually generated from the model.

The inferences on the number of stars $N$ and the luminosity function parameters
are shown in Figure~\ref{fig:results2}. The uncertainty about the luminosity
function parameters is still considerable despite the larger number of stars, as the fluxes of the fainter stars
are not well constrained by the data. The true values are still well within
the range of plausible values in the posterior distribution.
As with Test Case 1, the PSF parameters $\{s_1, s_2, w\}$ and the noise parameters $\{\sigma_0, \eta\}$ were also inferred accurately with small uncertainties.

\section{Comparison to SExtractor}\label{sec:sex}
In the previous section we established that the inference of the catalogs from
the data is computationally feasible and that the number of stars and the
luminosity function can be inferred from the image data, albeit with moderate
uncertainties. We now compare this approach to an alternative analysis that makes
use of the standard
tool \sex~\citep{sextractor}. To achieve this, we executed \sex~on the two test images, for various
values of the detection threshold parameters {\tt DETECT\_THRESH} and {\tt ANALYSIS\_THRESH}
ranging from 0.5 to 6.5. This results in a set of catalogs
for each image, with more conservative thresholds resulting in less stars detected
as compared to more aggressive thresholds. To compute flux
estimates that are directly comparable to the fluxes in our input catalogs, we
configured \sex~to compute object fluxes within fixed circular apertures that were
known to contain 70\% of the mass of the PSF. The \sex~flux estimates were then
scaled up to account for this finite aperture.

\begin{figure}[ht!]
\begin{center}
\includegraphics[scale=0.5]{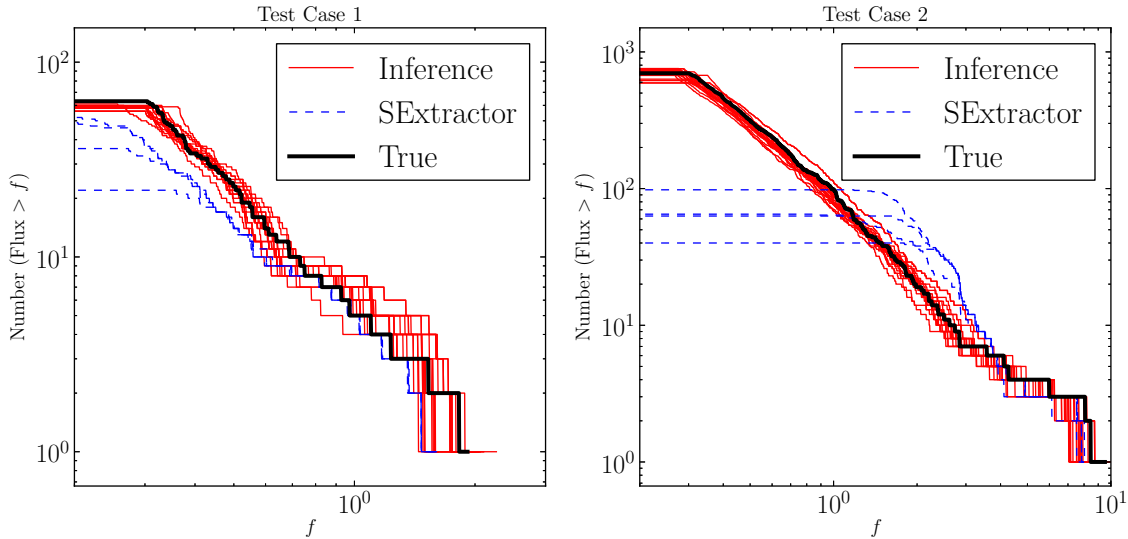}
\end{center}
\caption{The cumulative luminosity functions (number of stars above a given
flux, as a function of flux) produced by the Bayesian method
(several posterior samples shown)
and
\sex~(for various values of the threshold parameters), compared with the
actual cumulative LF. Both methods correctly identify the fluxes at the bright
end, with some uncertainty due to overlapping sources. However, at the lower
end \sex~is unable to detect all of the stars whereas the true CLF is typical
of the posterior distribution.
\label{fig:luminosity_function}}
\end{figure}

In Figure~\ref{fig:luminosity_function}, we present the cumulative luminosity
function (CLF) of the stars in the two fields, defined as the number of stars above
a given flux. The true CLF is plotted along with several
posterior samples from the Bayesian method and catalogs produced by \sex~for
various values of the detection and analysis thresholds.
We note that the true CLF is typical of the posterior samples, as expected.
For both test cases, the \sex~catalog is also consistent with the posterior
distribution at the bright end. However, the inferences from \sex~and the Bayesian
method differ at the faint end, with the former significantly underestimating
the number of faint stars.

This result may be attributable to the fact that the Bayesian method knows
about the existence and form of the luminosity function, even though it
does not know the values of the parameters. To test this, we ran the inference
on the data using an incorrect exponential distribution for the luminosity
function. The resulting CLF from this run did undershoot the true CLF
at the faint end.
However, the
Bayesian evidence for the exponential model was significantly lower (by a factor of
approximately $10^7$) than
for the (correct) broken power-law model.

In practice, we note that the wings of the PSF might become degenerate
with an nonzero flat background level in the data. To test whether this
was influencing the inferences (particularly about the faint end of the CLF)
we also ran a model that included an unknown constant background. This
had only a minor effect on the resulting inferences.

\section{Discussion and Conclusions}\label{sec:conclusion}

In this paper we have developed and demonstrated a Bayesian approach to
making catalogs from astronomical images in the case where the image contains
only stars (or other point sources). The key idea is that instead of computing
a single catalog, the method creates a posterior probability distribution on
the space of possible catalogs that represents our state of knowledge about
the presence and properties of objects in the image. When this is done, the
uncertainties in the imaging are accurately propagated through to scientific
conclusions, for example about the luminosity function of the stars.
This approach was contrasted with the results from the standard \sex~software.
For the bright sources the results were essentially consistent, however the
Bayesian approach was more successful at modelling the distribution of faint
stars. Of course, the Bayesian method is much more computationally intensive,
which is a significant issue in practice.
However, the great value of upcoming imaging data sets and the irreproducibility of astronomical imaging data in areas of time-domain astrophysics (for example in observations of rare events) make it important to extract as much information as possible from every patch of imaging.  Our view is that the additional CPU time and the non-triviality of the outputs from our method will be worth the effort in the next generation of astronomical experiments.
To build a more complete picture of
when our approach is necessary in practice, it will need to be tested against
a wider variety of alternative methods  such as DAOPHOT \citep{1987PASP...99..191S}
and DOLPHOT \citep{dolphot}.

We note that there are many limitations to the model presented in this paper,
some of which will be important to relax when it is applied to real data.
In principle, our model should be a model of the physical state of the universe,
and not a simple model where the only stellar properties are a 2-D position and
a flux. Another limitation is that we have not considered multi-epoch or multi-band
imaging. In the former case, PSF variations and stellar motions may be relevant
\citep{lang}, and in the latter, a model for the spectral energy distributions
of the stars will need to be considered: essentially, the luminosity function
will need to be a probability distribution over more than one dimension.

In practice, it may also be necessary to improve the model for the prior
distribution
of stellar positions and fluxes. One area where this is clearly needed is the
application of this approach to images of stellar clusters. The model would need
to be revised to take into account the fact that we expect the stars' positions
to be clustered together, whereas the current model implies a large prior
probability for the stars being scattered evenly across the image. In this and
other applications, the luminosity function would also require multiple
components, for example consisting of stars that are associated with a cluster
or a stream and those that are not.

Throughout this paper, we have also assumed that the pixel-convolved PSF can
be adequately modeled using simple components and that there are no PSF
variations across the field. Relaxing these assumption provides a significant
challenge for the future.



\section{Acknowledgements}
It is a pleasure to thank
  Anna Nierenberg (UCSB),
  Jonathan Goodman (NYU),
  Fengji Hou (NYU),
  Dustin Lang (CMU),
  Andrew Dolphin (Raytheon),
  Julianne Dalcanton (Washington),
  Geraint Lewis (Sydney), and
  Phil Marshall (Oxford)
for their comments and discussions.
The referee is thanked for their thoughtful comments which helped us to improve
the paper.
BJB would like to thank Tommaso Treu (UCSB) for his support, and
Wayne Stewart and Arden Miller (Auckland) for their encouragement and advice.
DFM and DWH were partially supported by
   the US National Science Foundation (grant IIS-1124794) and
   the US National Aeronautics and Space Administration (grant NNX12AI50G).

\appendix

\section{Broken Power-Law Distribution}\label{power_law}
\begin{figure}[ht!]
\begin{center}
\includegraphics[width=\textwidth]{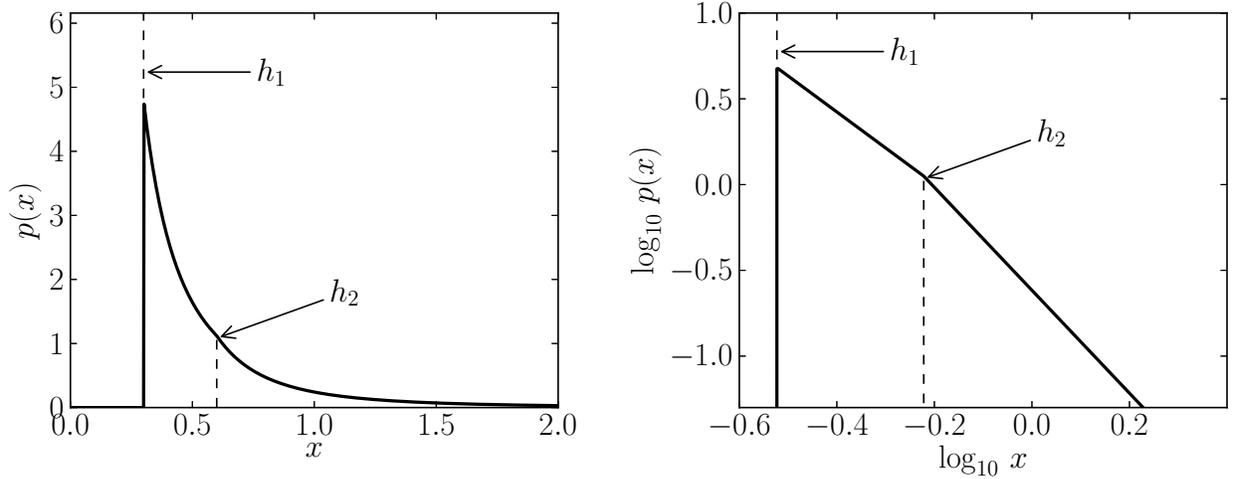}
\caption{A broken power-law distribution. The parameter values for this
particular PDF were
$\{h_1, h_2, \alpha_1, \alpha_2\} = \{0.3, 0.6, 1.1, 2\}$, i.e. the same
parameter values used to make the simulated data.
\label{fig:powerlaw}}
\end{center}
\end{figure}

The broken power-law distribution is based on a straightforward extension
to a simple power-law distribution (also known as a Pareto distribution,
particularly in the statistics literature). The power-law distribution for a
variable $x$ (given a lower cutoff $x=h$ and a slope $\alpha$) is defined by:
\begin{eqnarray}
p(x) &\propto&
\left\{
\begin{array}{lcl}
0, & & x < h \\
x^{-\alpha - 1} & & x \geq h.
\end{array}
\right.
\end{eqnarray}
In contrast, the broken power-law distribution for a variable $x$ is defined by
a lower cutoff $x=h_1$, two slopes $\{\alpha_1, \alpha_2\}$ and a break point
$x=h_2$:
\begin{eqnarray}
p(x) &\propto&
\left\{
\begin{array}{lcrl}
0, & & x < h_1 & \\
x^{-\alpha_1 - 1} & & h_1 \leq x \leq h_2 & \\
x^{-\alpha_2 - 1} & & x > h_2.
\end{array}
\right.
\end{eqnarray}
The free parameters of the broken power-law are:
\begin{eqnarray}
\beta = \{h_1, h_2, \alpha_1, \alpha_2\}.
\end{eqnarray}
With normalising terms included, the proportionality becomes an equality:
\begin{eqnarray}
p(x) &=&
\left\{
\begin{array}{lcr}
0, & & x < h_1 \\
Z_1^{-1}x^{-\alpha_1 - 1}, & & h_1 \leq x \leq h_2 \\
Z_2^{-1}x^{-\alpha_2 - 1}, & & x > h_2.
\end{array}
\right.
\end{eqnarray}
Two conditions will be used to determine the normalizers $Z_1$ and $Z_2$.
Firstly, the probability density function (PDF) should be continuous at $x=h_2$:
\begin{eqnarray}
Z_1^{-1}h_2^{-\alpha_1 - 1} &=& Z_2^{-1}h_2^{-\alpha_2 - 1}\\
\implies
Z_2 &=& Z_1h_2^{\alpha_1-\alpha_2}
\end{eqnarray}
The second condition is that the total probability must be 1:
\begin{eqnarray}
\int_{h_1}^{h_2} Z_1^{-1} x^{-\alpha_1 - 1} \, dx
+
\int_{h_2}^\infty Z_2^{-1} x^{-\alpha_2 - 1} \, dx
&=& 1 \\
Z_1^{-1}\alpha_1^{-1}\left[h_1^{-\alpha_1} - h_2^{-\alpha_1}\right]
+
Z_2^{-1}\alpha_2^{-1}h_2^{-\alpha_2}
&=& 1 \\
Z_1^{-1}\alpha_1^{-1}\left[h_1^{-\alpha_1} - h_2^{-\alpha_1}\right]
+
Z_1^{-1}h_2^{\alpha_2-\alpha_1}\alpha_2^{-1}h_2^{-\alpha_2}
&=& 1
\end{eqnarray}
\begin{eqnarray}
\implies
Z_1 &=& \alpha_1^{-1}\left[h_1^{-\alpha_1} - h_2^{-\alpha_1}\right]
+
h_2^{-\alpha_1}\alpha_2^{-1}.
\end{eqnarray}
The cumulative distribution (CDF) is a useful property of a probability
distribution and is given by the antiderivative of the PDF:
\begin{eqnarray}
P(X \leq x) = F(x) &=&
\left\{
\begin{array}{lcr}
0, & & x < h_1 \\
(Z_1\alpha_1)^{-1}\left(h_1^{-\alpha_1} - x^{-\alpha_1}\right), & & h_1 \leq x \leq h_2 \\
1 - (Z_2\alpha_2)^{-1}x^{-\alpha_2}, & & x > h_2.
\end{array}
\right.
\end{eqnarray}
The inverse of the CDF is also useful and is given by:
\begin{eqnarray}
F^{-1}(u) &=&
\left\{
\begin{array}{lcr}
\left[h_1^{-\alpha_1} - uZ_1\alpha_1\right]^{-1/\alpha_1}, & & 0 < u < 1 - (Z_2\alpha_2)^{-1}h_2^{-\alpha_2}\\
\left[Z_2\alpha_2(1-u)\right]^{-1/\alpha_2},& & 1 - (Z_2\alpha_2)^{-1}h_2^{-\alpha_2}< u < 1.
\end{array}
\right.
\end{eqnarray}
An example of a broken power-law distribution is shown in Figure~\ref{fig:powerlaw}.

\section{Proposal Distributions}
To implement Metropolis-Hastings moves for the space of possible catalogs,
proposal distributions are required. See Table~\ref{tab:proposals} for a list of
proposal distributions used in this study.

\begin{table}[ht!]\footnotesize
\begin{center}
\begin{tabular}{|c|c|c|}
\hline
Parameter & Proposal & Notes\\
\hline
$N$ & $N \to N + \delta_N$ & Generate $\delta_N$ new stars from
$p(x, y, f|\beta)$.\\
$N$ & $N \to N - \delta_N$ & Remove $\delta_N$ stars, chosen at random\\
$\beta$ & $\beta \to \beta + \delta_\beta$ & Transform
stars' fluxes correspondingly\\
$\beta$ & $\beta \to \beta + \delta_\beta$ & Fix stars' fluxes, put extra
term in acceptance probability \\
$(x_i,y_i)$ & $(x_i,y_i) \to (x_i,y_i)+(\delta_x, \delta_y)$ & Can move $>1$ star
in a single step \\
$f$ & $f \to f + \delta_f$ & Can move $>1$ stars' fluxes in a single step\\
\hline
\end{tabular}
\end{center}
\caption{All $\delta$ parameters are drawn from multi-scale distibutions such
that the largest steps are of order the prior width, and the smallest steps
are of order $10^{-6}$ times the prior width.\label{tab:proposals}}
\end{table}



\end{document}